\documentclass[12pt]{article}
\usepackage{amssymb}
\usepackage{amsmath}
\def\dd{{\rm d}}\def\ee{{\rm e}}\def\ii{{\rm i}}
\def\SU{{\rm SU}}
\def\half{{\textstyle{\frac12}}}\def\fourth{{\textstyle{\frac14}}}
\def\beq{\begin{equation}}\def\eeq{\end{equation}}
\def\bea{\begin{eqnarray}}\def\eea{\end{eqnarray}}

\begin{document}
\title{Gauge Fields in Causal Set Theory}
\author{Roman Sverdlov
\\Physics Department, University of Michigan,
\\450 Church Street, Ann Arbor, MI 48109-1040, USA}
\date{July 13, 2008}
\maketitle

\begin{abstract}
\noindent This is the second paper in a series on the dynamics of matter fields in the causal set approach to quantum gravity. We start with the usual expression for the Lagrangian of a charged scalar field coupled to a SU($n$) Yang-Mills field, in which the gauge field is represented by a connection form, and show how to write it in terms of holonomies between pairs of points, causal relations, and volumes or timelike distances, all of which have a natural correspondence in the causal set context. In the second part of the paper we present an alternative model, in which the gauge field appears as the result of a procedure inspired by the Kaluza-Klein reduction in continuum field theory, and the dynamics can be derived simply using the gravitational Lagrangian of the theory.
\end{abstract}

\noindent{\bf 1. Introduction}

\noindent The work described in this paper is a continuation of that in Ref \cite{paperI}, and is part of a program to recast the Lagrangian for various types of fields in terms appropriate to the causal set approach to quantum gravity. A causal set \cite{causets1,causets2} is a locally finite partially ordered set, where the partial order is interpreted as giving the causal relations among its elements. A causal set is considered to be ``manifold-like" if its elements can be embedded with uniform density in a Lorentzian manifold, in such a way that the causal relations are preserved. The idea behind the causal set approach to quantum gravity is that, when such an embedding exists, the whole structure of the Lorentzian manifold at scales larger than the embedding density is in fact encoded in the purely combinatorial structure of the causal set.

The fields considered in this paper are a SU($n$) gauge field and a charged spin-0 field interacting with the gauge field, both defined on a causal set. In order to formulate the dynamics of these fields, we will rewrite the action for each of them in terms of variables which only require, as structures related to the underlying set, the notions of volumes and causal relations. We start by doing this for the charged scalar field in the next section, which essentially just involves adapting the results for uncharged scalar fields of Ref \cite{paperI}, and then derive the causal set version of the Yang-Mills action for the gauge field in Sec 3. Sec 4 contains the causal set version of an alternative, Kaluza-Klein-type approach to an (Abelian) gauge theory coupled to gravity, and Sec 5 contains concluding remarks.

\bigskip
\noindent{\bf 2. Causal Set Discretization of the Charged Scalar Field Lagrangian}

\noindent In this section, we consider a charged spin-0 particle, described by a set of complex scalar fields $\phi = (\phi_1, ..., \phi_n)$ coupled to a SU($n$) gauge field. (Notice that in this approach to the dynamics of matter fields in causal set theory, although I will assume that spacetime is discretized, the internal degrees of freedom will still have a continuous invariance group.) The dynamics of such a field can be described in the continuum starting with the matter Lagrangian density
\beq
{\cal L}_{\rm m}(g_{\mu\nu},\phi,A_\mu;x) = \half\, |g|^{1/2}\,\big[g^{\mu\nu}\,
(D_\mu\phi)^{\dagger}\,(D_\nu\phi) - m^2\,\phi^{\dagger}\phi\big]\;, \label{Lm}
\eeq
where the gauge covariant derivative is defined as usual by
$D_\mu\phi^a:= \partial_\mu\phi^a + \ii\,e\, A_\mu{}^a{}_b\,\phi^b$,
with $A_\mu = A_\mu{}^k\,T^k$ the Lie-algebra-valued connection form representing the gauge field on a differentiable manifold. (Here, Latin indices $a$, $b$, ..., are Lie-algebra tensor indices, while $k$, $l$, ..., label elements of the basis $T^k$ of the Lie algebra.) In the causal set context, the scalar field will be simply replaced by a corresponding field defined at each causal set element, but to write down the action we need to specify what variables will replace $A_\mu$.

We would want to redefine the gauge field without referring to the differentiable structure of the manifold, as implied by the presence of the spacetime tensor index $\mu$. The most natural quantity to use is the holonomy, the group transformation corresponding to the parallel transport of a Lie-algebra-valued field such as $\phi$ between two points $p$ and $q$. In a differentiable Lorentzian manifold $M$ (of dimension $d$), we define the holonomy as the function $f : M\times M \rightarrow \SU(n)$ which assigns to any two elements $p,\,q\in M$ the holonomy of $A_\mu$ along the geodesic segment $\gamma(p,q)$ connecting $p$ and $q$ in $M$, given by the path-ordered exponential
\beq
f(p,q) = {\rm P}\,\exp\Big\{\ii\,e\int_{\gamma(p,q)}
A_\mu{}^k\,T^k\,\dd x^\mu\Big\}\;, \label{holo}
\eeq
in terms of which the expression $D_\mu\phi(x)$ appearing in the scalar field Lagrangian arises from the leading-order term in the expansion of the expression $(f(x,y)\,\phi(y) - \phi(x))$ \cite{PS}.

This means that we can obtain the causal set version of the charged scalar field Lagrangian by making some simple substitutions in the one obtained in Ref \cite{paperI} for the Klein-Gordon field. Thus, for a causal set $(S,\prec)$ and matter fields $\phi$ and $f$ defined on $S$,
\bea
& &{\cal L}_{\rm m}(\prec,\tau,\phi,f;p,q)\\
& &= \bigg(1+\frac{I_{d,0}+\fourth\,k_d}{I_{d,1}} \bigg)\,
\frac{(\phi(q)-\phi(p))^2}{2\,\tau^2(p,q)} - \frac{\rho^{-1}}{2\,I_{d,1}\,
\tau^{d+2}(p,q)}\sum_{p\prec x\prec q}(\phi(x)-\phi(p))^2
- \half\,m^2\,\phi(p)^2\;, \label{KGd}\nonumber
\eea
where, for manifold-like causal sets, $\rho$ is interpreted as the density of the embedding in the $d$-manifold $M$ and $\tau(p,q)$ is the Lorentzian distance from $p$ to $q$, which can be obtained from the length of the longest chain between those points, and the various constants in this expression are given by
\bea
& &I_{d,0} = \frac{c_{d-1}}{2^d\, d\, (d+1)\,(d+2)}\;, \qquad
I_{d,1} = \frac{c_{d-2}\,J_{d+1}}{2^{d+1}\,d\,(d+2)}\;, \qquad
J_d:= \int_{-\pi/2}^{\pi/2} (\cos\theta)^d\,\dd\theta\;, \qquad\nonumber\\
& &k_d:= \frac{c_{d-1}}{2^{d-1}\,d}\;, \qquad
c_d = \frac{2\,\pi^{d/2}}{d\,\Gamma(d/2)}\;. \label{const}
\eea
(The $c_d$ are the constants in terms of which the volume of the $d$-dimensional ball of radius $r$ in Euclidean space is $c_d\,r^d$, and the $k_d$ those in terms of which the volume of the $d$-dimensional Alexandrov set of height $\tau$ in Minkowski space is $k_d\,\tau^d$.) In the case of more general, non-manifoldlike causal sets, $d$ and $\rho$ are just a pair of parameters and, since timelike distances between points are not defined, we can make the expression (\ref{KGd}) meaningful by re-expressing $\tau(p,q)$ in terms of the volume $V(p,q)$ of the Alexandrov set $\alpha(p,q)$ of $p$ and $q$, using the relationship between then that would hold in the manifoldlike case \cite{paperI},
\beq
V(p,q) = \frac{c_{d-1}}{2^{d-1}\,d}\,\tau^d(p,q)\;.\label{V}
\eeq
As with the uncharged Klein-Gordon field, the Lagrangian in (\ref{KGd}) is a {\em quasilocal\/} one, whose expression in the continuum tends to the usual one as $q\to p$ in the manifold topology.

\bigskip
\noindent{\bf 3. Causal Set Discretization of the Yang-Mills Lagrangian}

\noindent In this section, our main goal is to express the Yang-Mills Lagrangian density,
\beq
{\cal L}_{\rm YM}(g_{\mu\nu},A_\mu;x)
= \half\,|g|^{1/2}\,{\rm tr}(F_{\mu\nu}F^{\mu\nu})\;,
\eeq
in terms of the holonomy variables for the gauge field introduced in the previous section, as well as variables describing the geometry that are meaningful in the causal set context, namely causal relations, and either volumes or timelike lengths. Once this is done, the Lagrangian density can be easily rewritten in the discrete setting.

To begin, let us recall the relationship between the holonomies (\ref{holo}) and the curvature $F_{\mu\nu}{}^k$ of $A_\mu{}^k$ that appears in the Lagrangian density. Consider a region small enough for $F_{\mu\nu}{}^k$ to be approximately constant, and pick three points in that region, $a$, $b$, and $c$. In this region we can approximate spacetime with a portion of Minkowski space, and assume for definiteness that the three points are spacelike related. Choose a coordinate system so that $a$ coincides with the origin, the $x$ axis points from $a$ to $b$, and the $y$ axis is perpendicular to the $x$ axis in the $abc$ plane. Then in this coordinate system $a = (0,0,0,...)$, $b = (0,b^1,0,...)$, $c = (0,c^1,c^2,...)$. The fact that the contour integral of $A_\mu{}^k$ around the triangle $abc$ is equal to the flux of $F_{\mu\nu}{}^k$ through the interior of that triangle is expressed by the relationship
\beq
f(a,b)\, f(b,c)\, f(c,a) = 1 + \half\, b^1\, c^2\, F_{12}{}^k\,T^k + ...
\eeq
This result generalizes to points at arbitrary locations, and can be written covariantly as
\beq
f(a,b)\, f(b,c)\, f(c,a) = 1 + \half\, F_{\mu\nu}{}^k\,T^k
(b^\mu - a^\mu)(c^\nu - a^\nu) + ...
\eeq
Recalling that, for SU($n$), tr$(T^kT^l) = C_2\, \delta_{kl}$, we get that, to leading order in the separation between points,
\beq
{\rm tr}[(f(a,b)f(b,c)f(c,a)-1)^2] = \frac{C_2}{4} F_{\mu\nu}{}^k (b^\mu - a^\mu)(c^\nu - a^\nu)F_{\rho\sigma}{}^k\, (b^\rho - a^\rho)(c^\sigma - a^\sigma)\;.
\label{trabc}
\eeq
The left-hand side of this equation does not make any reference to tensor indices, thus it is a good building block to express the Lagrangian for a gauge field on a causal set. 

I am going to proceed in a similar fashion to the way in which we constructed an action for a scalar field on a causal set in Ref \cite{paperI}: I will consider an Alexandrov set $\alpha(p,q)$ based on two points $p \prec q$, and obtain a system of two equations; one equation will come from integrating the above expression over all possible $a$, $b$ and $c$ inside $\alpha(p,q)$, the other one will come from integrating only over $c$, while setting $a = p$ and $b = q$. If we fix the coordinate system so that $p = (-\frac{\tau}{2},0,0,0)$ and $q = (\frac{\tau}{2},0,0,0)$, where $\tau$ is the proper time between $p$ and $q$, the two unknowns will end up being $F_{0i}{}^k\, F_{0i}{}^k$ and  $F_{ij}{}^k\,F_{ij}{}^k$; solving the system of equations and subtracting one unknown from the other will then give us $F^{\mu\nu\,k}\,F_{\mu\nu}{}^k$.

I will start from the case in which all three points are allowed to move throughout the Alexandrov set, expand the right-hand side of Eq (\ref{trabc}), and integrate term by term. Clearly any term with an odd number of powers of any variable will integrate to 0. Thus, the only terms that may potentially survive the integration are those of the form $a^{\mu}\, a^{\nu}\, a^{\rho}\, a^{\sigma}$ or quadratic terms in two of the three points. By a simple counting of terms we get
\bea
& &\int_{p\prec a,b,c\prec q} \dd^da\,\dd^db\,\dd^dc\; {\rm tr}\big[(f(a,b)\,f(b,c)\,f(c,a)-1)^2\big] \nonumber \\
& &= \frac{C_2}{4}\, F_{\mu\nu}{}^k\, F_{\rho\sigma}{}^k \int_{p\prec a,b,c\prec q} \dd^da\, \dd^db\,\dd^dc\; (b^\mu - a^\mu)(c^\nu - a^\nu)(b^\rho - a^\rho)(c^\sigma - a^\sigma) \nonumber \\
& &= \frac{C_2}{4}\; \bigg[ 3\,V\sum_{k,\mu,\nu} (F_{\mu\nu}{}^k)^2
\bigg(\int_{\alpha(p,q)} \dd^da\, (a^\mu)^2 \bigg) \bigg(\int_{\alpha(p,q)} \dd^db\, (b^\nu)^2 \bigg) \nonumber \\
& & \kern60pt -\ V^2 F_{\mu\nu}{}^k\, F_{\rho\sigma}{}^k\int_{\alpha(p,q)}
\dd^da\, a^\mu\, a^\nu\, a^\rho\, a^\sigma \bigg]\;, \label{intabc}
\eea
where $V$ is the volume of the Alexandrov set $\alpha(p,q)$.

The only terms of $F_{\mu\nu}{}^k\, F_{\rho\sigma}{}^k\, a^{\mu}\, a^{\nu} a^{\rho} a^{\sigma}$ that survive integration are the ones whose indices are pairwise equal. But if either $\mu = \nu$  or $\rho = \sigma$  then we get $F_{\mu\nu}{}^k = 0$ or $F_{\rho\sigma}{}^k = 0$, respectively, which would set the whole thing to 0. Thus, our only options are $\mu = \rho$, $\nu = \sigma$ and $\nu = \rho$, $\mu = \sigma$.  The antisymmetry of $F_{\mu\nu}{}^k$ then implies that these two cases are opposites of each other, which in turn implies that $F_{\mu\nu}{}^k\, F_{\rho\sigma}{}^k\,a^\mu\, a^\nu\, a^\rho\, a^\sigma = 0$. Thus, Eq (\ref{intabc}) becomes
\bea
& &\int_{p\prec a,b,c\prec q} \dd^da\,\dd^db\,\dd^dc\;
{\rm tr} \big[ (f(a,b)\,f(b,c)\,f(c,a)-1)^2 \big]
\nonumber\\
& &= \frac{3\,VC_2}{4} \sum_{k,\mu,\nu} (F_{\mu\nu}{}^k)^2 \Big(\int_{\alpha(p,q)}
\dd^da\, (a^\mu)^2\Big) \Big(\int_{\alpha(p,q)} \dd^db\, (b^\nu)^2\Big) \nonumber\\
& &= \frac{3\,VC_2}{2}\, \sum_k \Big(J^0 J^1 \sum_{i=1}^{d-1}
(F_{i0}{}^k)^2 + (J^1)^2 \sum_{i<j} (F_{ij}{}^k)^2 \Big)\;, \label{intabc2}
\eea
where $J^{\mu}=\int_{\alpha(p,q)} \dd^dx\, (x^\mu)^2$, or in other words
\beq
J^0 = I_{d,0}\,\tau^{d+2}
= \frac{c_{d-1}\,\tau^{d+2}}{2^d\, d\,(d+1)\,(d+2)}\;,\quad
J^1 = ... = J^{d-1} = I_{d,1}\,\tau^{d+2}
= \frac{c_{d-2}\,J_{d+1}\, \tau^{d+2}}{2^{d+1}\, d\,(d+2)}\;.
\eeq
We thus have two unknowns, $\sum_{i=1}^{d-1} (F_{i0}{}^k)^2$ and $\sum_{i<j} (F_{ij}{}^k)^2$, and we need one more equation. I will get my second equation by evaluating $\int_{\alpha(p,q)} \dd^dx\;{\rm tr}[(f(p,x)\,f(x,q)\,f(q,p)-1)^2]$, where $p \prec q$ are the endpoints of the Alexandrov set. If we rewrite Eq (\ref{trabc}) in terms of the points $p$, $x$, and $q$, we get
\bea
& &{\rm tr}\big[ (f(p,x)\,f(x,q)\,f(q,p)-1)^2 \big] \\ \noalign{\medskip}
& &=\ \fourth\,C_2\,F_{\mu\nu}{}^k\,(p^\mu-x^\mu)\,(q^\nu-x^\nu)\,F_{\rho\sigma}{}^k\, (p^\rho-x^\rho)\,(q^\sigma-x^\sigma)\;. \nonumber
\label{trApB}
\eea
Again we can expand it and integrate term by term. There are several conditions each term has to meet, in order for its integral not to vanish. First of all, it needs to contain an even number of factors of $x$. Secondly, as we have seen before, for symmetry reasons
\beq
F_{\mu\nu}{}^k\, F_{\rho\sigma}{}^k \int_{\alpha(p,q)}\, \dd^dx\,
\, x^\mu\, x^\nu\, x^\rho\, x^\sigma = 0\;.
\eeq
Finally, $F_{\mu\nu}\, p^\mu p^\nu = F_{\mu\nu}\, q^\mu q^\nu = 0$
and, since $p = (-\frac{\tau}{2},0,0,0)$ and $q = (\frac{\tau}{2},0,0,0)$, we have $F_{\mu\nu}\, p^\mu q^\nu = -F_{\mu\nu}\, p^\mu p^\nu = 0$. The only terms in Eq (\ref{trApB}) that do {\em not\/} vanish for any of the above reasons are 
\bea
& &F_{\mu\nu}{}^k\, F_{\rho\sigma}{}^k\, p^\mu\, x^\nu\, p^\rho\, x^\sigma\,,\qquad
F_{\mu\nu}{}^k\, F_{\rho\sigma}{}^k\, p^\mu\, x^\nu\, x^\rho\, q^\sigma\,,
\nonumber\\ \noalign{\medskip}
& &F_{\mu\nu}{}^k\, F_{\rho\sigma}{}^k\, x^\mu\, q^\nu\, p^\rho\, x^\sigma\,,\qquad
F_{\mu\nu}{}^k\, F_{\rho\sigma}{}^k\, x^\mu\, q^\nu\, x^\rho\, q^\sigma\,.
\nonumber
\eea
Plugging in the coordinate values of $p$ and $q$ we see that each
of the above four expressions evaluates to 
$\fourth\,\tau^2\, F_{\mu0}{}^k\, F_{\rho0}{}^k\, x^\mu\, x^\rho$.
In order for this not to be an odd function we need $\mu = \rho$, and
in order for $F_{\mu 0}$ to be non-zero we need $\mu \not= 0$. Thus,
this becomes $\fourth\,\tau^2\, (F_{i0}{}^k)^2\, (x^i)^2$ and, since
there are four such terms, the integral becomes
\bea
&& \int_{\alpha(p,q)} \dd^dx\;{\rm tr}\big[ (f(p,x)\,f(x,q)\,f(q,p)-1)^2 \big]
\nonumber \\
&&= \fourth\,C_2\, \tau^2 \sum_{i=1}^{d-1} \int_{\alpha(p,q)} \dd^dx\,(F_{i0}{}^k)^2\, (x^i)^2 = \fourth\,C_2\, \tau^2 J^1 \sum_{i=1}^{d-1} (F_{i0}{}^k)^2\;,
\eea
where I have used the fact that $J^1 = ... = J^{d-1}$, from rotational symmetry.
We thus get
\beq
\sum_{i=1}^{d-1} (F_{i0}{}^k)^2 = \frac{4}{C_2\, \tau^2 J^1}
\int_{\alpha(p,q)} \,\dd^dx\; {\rm tr}\big[ (f(p,x)\,f(x,q)\,f(q,p)-1)^2 \big]\;,
\label{sumFi02}
\eeq
and if we recall Eq (\ref{intabc2}), we similarly get
\bea
& &\sum_{i<j} (F_{ij}{}^k)^2 = \frac{4}{C_2\, (J^1)^2} \bigg(\frac{1}{6V}
\int_{p\prec a,b,c\prec q} \dd^da\, \dd^db\, \dd^dc\;
{\rm tr} \big[ (f(a,b)\,f(b,c)\,f(c,a)-1)^2 \big]\nonumber\\
& &\kern115pt -\ \frac{J^0}{\tau^2} \int_{\alpha(p,q)} \dd^dx\;{\rm tr}\big[ (f(p,x)\,f(x,q)\,f(q,p)-1)^2 \big] \bigg)\;.\ \label{sumFij2}
\eea
Finally, plugging the expressions we obtained for $J^0$, $J^1$ and $V$ into (\ref{sumFi02}) and (\ref{sumFij2}) we get
\bea
& &{\rm tr}\,(F^{\mu\nu}F_{\mu\nu}) \nonumber\\
& &= 2\,\Big(\sum_{i=1}^{d-1} (F_{i0}{}^k)^2
+ \sum_{i<j} (F_{ij}{}^k)^2 \Big) \nonumber\\
& &= \frac{1}{C_2}\, \bigg[\frac{d^3\, (d+2)^2\, 8^{d+1}}{6\,c_{d-1}\,c_{d-2}^2
\,J_{d+1}^2\,\tau^{3d+4}} \int_{p\prec a,b,c\prec q} \dd^da\, \dd^db\, \dd^dc\; {\rm tr}\big[ (f(a,b)\,f(b,c)\,f(c,a)-1)^2 \big] \nonumber\\
& &\kern36pt-\ \frac{2^{d+3} d\,(d+2)}{c_{d-2}\, J_{d+1}\,\tau^{d+4}}
\Big(1+\frac{2\,c_{d-1}}{c_{d-2}\,J_{d+1}\,(d+1)}\Big) \times\ \nonumber\\
& &\kern80pt \times\ \int_{\alpha(p,q)} \,\dd^dx\;
{\rm tr}\big[ (f(p,x)\,f(x,q)\,f(q,p)-1)^2 \big] \bigg]\;.
\eea
Alternatively, we can rewrite the above in terms of $V$ instead of $\tau$, as follows:
\bea
& &{\rm tr}\,(F^{\mu\nu}F_{\mu\nu}) \nonumber\\
& &= \frac{1}{C_2}\, \bigg[\frac{d^3\, (d+2)^2\, 8^{d+1}}{6\,c_{d-1}\,c_{d-2}^2\, J_{d+1}^2} \Big(\frac{c_{d-1}}{2^{d-1}\,d\,V}\Big)^{3+4/d} \int_{p\prec a,b,c\prec q} \dd^da\, \dd^db\, \dd^dc\;{\rm tr}\big[ (f(a,b)\,f(b,c)\,f(c,a)-1)^2 \big] \ \nonumber\\
& &\kern36pt-\ \frac{2^{d+3}\, d\,(d+2)}{c_{d-2}\, J_{d+1}} \Big(\frac{c_{d-1}}{2^{d-1}\,d\,V}\Big)^{1+4/d} \Big(1+\frac{2\,c_{d-1}}{c_{d-2}\,J_{d+1}\,(d+1)}\Big) \times\ \nonumber\\
& &\kern80pt\times \int_{\alpha(p,q)} \,\dd^dx\;{\rm tr}\big[ (f(p,x)\,f(x,q)\,f(q,p)-1)^2 \big] \bigg]\;.
\eea
Again, both of these expressions depend on two points $p$ and $q$, and are to be used to construct a quasilocal Lagrangian density ${\cal L}_{\rm YM}(\prec,\tau,f;p,q) = \half\,{\rm tr}(F^{\mu\nu}F_{\mu\nu})(p,q)$.

\bigskip
\noindent{\bf 4. Kaluza-Klein Theory}

\noindent We have now completed the description of bosonic fields on causal sets in terms of holonomies. We will now shift gears and introduce a Kaluza-Klein-like model for a gauge field on a causal set. This will give us the option of choosing whether to view a gauge field as an independent entity from the other degrees of freedom, as presented in the picture with holonomies, or to view it as part of the gravitational field, as one does in the Kaluza-Klein model.

We recall that the Lagrangian for the gravitational theory, as obtained in Ref \cite{paperI}, can be written as
\beq
R = \frac1D\, \bigg\{ \bigg(\frac{V(\tau)}{k_d\,\tau^d} - 1\bigg)\,
\big(I_{d,0} + I_{d,1} + \fourth\,k_d\big)\,\tau^d
- \int_{\alpha(p,q)} \bigg(\frac{V(p,x)}{k_d\,\tau^d(p,x)} - 1\bigg)\,\dd^dx
\bigg\}\,. \label{grav}
\eeq
where
\beq
D = \bigg(\frac{d^2}{24\,(d+1)\,(d+2)}
+ \frac{d\,I_{d,1}}{24\,(d+1)} \bigg)\, \tau^{d+2}\;. \eeq
However, in a Kaluza-Klein-type approach, while we will still use the above equation, we would like to re-think what we mean by causal relations, volumes, etc., in order to know what to ``plug into" that equation. For simplicity, we will restrict ourselves to a situation that can be modeled after a U(1) Kaluza-Klein theory, with only one ``extra dimension".

The first issue is that even outside of the Kaluza-Klein context, how to make causal sets manifoldlike is a very problematic question. Therefore, the intention of this paper was to write a general Lagrangian, without reference to a manifold structure, that simply happens to coincide with the expected Lagrangian in the special case of manifoldlike causal sets. However, in the case of Kaluza-Klein theory, in order to obtain an effective four-dimensional theory, not only do we have to assume the manifold structure, but we have to go so far as to assume translational symmetry along the fifth dimension. This rotational symmetry contradicts the letter and spirit of causal sets, since the causal relations between different points on that fifth dimensional circle are supposed to be independent of each other, just like it is the case for any other arbitrary points in  a causal set. 

The way I propose to resolve this problem is to formally separate the ``extra dimension" from the rest of the structure of the causal set, as follows: I will view the entire circle along the extra dimension, as opposed to a selected point on that circle, as an element of the causal set $S$. The actual spacetime will no longer be the set $S$, but rather $S \times G$, where $G$ is a unit circle in the complex plane. The set $S$ is equipped with a partial order $\prec$, whose physical meaning is as follows: if $p$ and $q$ are two circles, then $p \prec q$ if and only if we can select at least one point on $p$ and at least one point on $q$ in such a way that they are causally related. We now need to introduce a different partial order, in order to define what we mean by the word ``causally related" in the last sentence. To do so, we introduce two real-valued functions, $d \colon S \times S \to \mathbb{R}$ and $f \colon S \times S \to \mathbb{R}$. We will then define a partial order $\prec_{d,f}$ on $S  \times G$ defined as follows: 

\noindent{\em Definition:\/} Let $p $ and $q $ be two elements of $S$ and let $c_1$ and $c_2$ be two elements of $G$. We say that $(p, c_1) \prec_{d,f} (q, c_2)$ if and only if there is a real number $r$ such that $c_2 = c_1\, \ee^{\ii r}$ satisfying
$$
-\half\,d(p,q) + f(p,q) \leq r \leq \half\,d(p,q) + f(p,q)\;.
$$
Thus, $d(p,q)$ is interpreted as the fraction of the circle $q$ that is covered by the light cone of an arbitrary point on the circle $p$, and $f(p,q)$ is interpreted as the off-center displacement of the portion of the circle $q$ covered by a point on the circle $p$. Thus, in manifold language, $d(p,q)$ is related to $g_{dd}$ (which in the Kaluza-Klein model is interpreted as a scalar field),\footnote{We are adopting the common convention in Kaluza-Klein theory of calling the extra dimension the $d$-th one, although $t$ is the 0-th dimension, so there is no ``$(d-1)$-th dimension").} while $f(p,q)$ is related to $g_{\mu d}$ (which is interpreted as a gauge field). One has to note that due to the fact that $d$ and $f$ are only functions of pairs of circles, and {\em not\/} of individual points on these circles, $(p,c_1) \prec_{d,f} (q, c_2)$ if and only if $(p, c_1 + c_3) \prec_{d,f} (q, c_2 + c_3)$, for any given $c_3 \in G$. This implies translational symmetry along each circle, as commonly required in a Kaluza-Klein model.

We would now like to impose constraints on $d$ and $f$ in such a way that the following two conditions are satisfied:

\noindent(1) Consistency between the two order relations: If $p$ and $q$ are two circles, then $p \prec q$ if and only if we can find at least one choice of $c_1$ and $c_2$ such that $(p, c_1)  \prec_{d,f} (q,c_2)$;

\noindent(2) Transitivity of the new order relation: If $p \prec_{d,f} r$ and $r \prec_{d,f} q$, then $p \prec_{d,f} q$.

\noindent It is easy to see that condition 1 can be enforced by the following requirement: $d(p,q) = 0$ if and only if neither $p \prec q$ nor $q \prec p$. Furthermore, for all $p$ and $q$, $d(p,q) = d(q,p)$.

Regarding condition 2, if the apply it to the ``right" edge of a circle, then $\half\,d+f$ will characterize the off-center displacement of that edge. This would give us the following inequality:
\beq
\half\,d(p,r) + f(p,r) + \half\,d(r,q) + f(r,q)
\leq \half\,d(p,q) + f(p,q)\;. \label{2a}
\eeq
If we now apply it to the ``left" edge, then $\half\,d-f$ will characterize the off-center displacement of the edge. This would give us the following: 
\beq
\half\,d(p,r) - f(p,r) + \half\,d(r,q) - f(r,q)
\leq \half\,d(p,q) - f(p,q)\;. \label{2b}
\eeq
If we add Eqs (\ref{2a}) and (\ref{2b}) we get
\beq
d(p,r)+d(r,q) \leq d(p,q)\;, \label{2primea}
\eeq
while if we subtract them we get
\beq
f(p,r)+f(r,q) \leq f(p,q)\;. \label{2primeb}
\eeq

We would now like to define volumes on $S \times G$; in particular, we are interested in being able to apply the definition to the calculation of the volume of an Alexandrov set, since that is what we have to use in (\ref{grav}). We define it as follows:

\noindent{\em Definition:\/} Let $T$ be a subset of $S \times G$. Then Vol$(T) = \sum_{p \in S} \mu (T \cap (\{p\} \times G))$, where $\mu$ is a measure taken from real analysis. 

Suppose we have two circles, $p$ and $q$, and on each of these circles we select a point with coordinate $r = 0$. Thus, we are looking at points $(p,1)$ and $(q,1)$. We would like to find the volume of the Alexandrov set $\alpha((p,1),(q,1))$:
$$
{\rm Vol}\big(\alpha((p,1),(q,1))\big)
= \sum_{p \prec r \prec q} \mu (r \cap \alpha ((p,1), (q,1)))\;,
$$
for which we need to calculate $\mu(r \cap \alpha ((p,1),(q,1))$.

The right edge of the part of $r$ that is causally related to $(p,1)$ is $(r, \ee^{\ii(d(p,r)/2+f(p,r))})$, and the right edge of the part of $r$ that is causally related to $q$ is $(r, \ee^{\ii(d(q,r)/2 + f(q,r))})$. Thus, the right-hand side of $r \cap \alpha((p,1),(q,1))$ is $(r, \ee^{\ii\,\min\{d(p,r)/2 + f(p,r), d(q,r)/2 + f(q,r)\}})$. On the other hand, the left edge of the part of $r$ that is causally related to $(p,1)$ is $(r, \ee^{\ii\,(-d(p,r)/2 + f(p,r))})$, and the left edge of the part of $r$ that is causally related to $q$ is $(r, \ee^{\ii\,(-d(q,r)/2 + f(q,r))})$. Thus, the left-hand side of $r \cap \alpha((p,1),(q,1))$ is $(r, \ee^{-\ii\,\min\{d(p,r)/2 - f(p,r), d(q,r)/2 - f(q,r)\}})$. This means that 
\bea
& &\mu(r \cap \alpha ((p,1),(q,1))
= \min\{\half\,d(p,r) - f(p,r), \half\,d(q,r) - f(q,r)\} \nonumber\\
& &\kern122pt+\ \min\{\half\,d(p,r) + f(p,r), \half\,d(q,r) + f(q,r)\} \;,
\eea
which gives us 
\bea
& &{\rm Vol}(\alpha((p,1),(q,1))
= \sum_{p \prec r \prec q} \big(\min\{\half\,d(p,r)-f(p,r), \half\,d(q,r)-f(q,r)\}
\nonumber\\
& &\kern132pt+\ \min\{\half\,d(p,r)+f(p,r), \half\,d(q,r)+f(q,r)\} \big) \;.
\eea

Despite the fact that the definition of volume is different, the definition of distance is the same as for regular causal sets. As we recall, in regular causal set theory, if $p \prec q$ then the distance between $p$ and $q$ is related to the length of the longest possible chain of points $(r_1, ..., r_n)$ satisfying $p \prec r_q \prec . . . \prec r_n \prec q$. Now, in the case of circles, it would be sufficient for only one point on one circle to be causally related to one point in the other circle in order for us to be able to draw a causal path through these circles. This means that we need $f(r_i,r_{i+1})>0$. But from the earlier discussion we know that, if $r$ and $s$ are not causally related, then $f(r,s)=0$ no matter how close they might be to each other's light cones. Thus, $f(r_i, r_{i+1})>0$ is equivalent to $r_i$ and $r_{i+1}$ being causally related, which means that if we start from the definition of distance based on $f$ we will end up with a definition of distance based on $\prec$.
 
One thing to note is that the range of values $f(p,q)$ can take is continuous. Thus, in our extra dimension the causal set is continuous, while in the other ones it is still discrete. By remembering that $f(p,q)$ is really just a scalar field, it is easy to see that this would not pose an infinity problem any more than the continuous gauge field did. Of course, however, we can always discretize it by hand, by restricting $f(p,q)$ to take on only values which are multiples of $1/N$, where $N$ is some large number. Neither discretizing nor failure to do so create any problems, so whether or not we want to do it is a question of aesthetics. On the one hand, a strong believer in Democritus might want to discretize the extra dimension for the sake of discreteness. Furthermore, even if one doesn't care about Democritus, it still doesn't seem logical that the dimensions are ``different" from each other. Thus, in order to make them the same we are forced to discretize the extra dimension. On the other hand, one can object and say that they prefer {\em not\/} to think of the ``extra dimension" as an ordinary dimension, but rather they would like to take seriously its interpretation of being just a field. This can be motivated by the question that if the extra dimension is no different from the other ones, what was the physical ``force" that made it into a circle and forced translational symmetry? Since I see the point in both of these arguments, I will leave it up to the reader to decide whether or not to discretize the fifth dimension.
 
We have to note that, if we knew that $g_{dd}$ is constant, we would also know that $d(p,q)$ is proportional to $\tau(p,q)$, if the circles are ``large" enough---or else it is always 1 or 0 if they are too small. This means that the inclusion of $d(p,q)$ would only be redundant. Thus, as stated earlier, the physical meaning of $d(p,q)$ is $g_{dd}$ or in other words the scalar field $\phi$. This means that when we perform path integrations, in order to evaluate $\int \dd g_{dd}\,(...)$, instead of writing $\int \dd(d(p,q))\, (...)$ we have to write $\int \dd(d(p,q))/\tau(p,q)\, (...)$, since $g_{dd} = d(p,q)/\tau(p,q)$. Apart from that, we also need to take into account the fact that, due to the $g_{\mu d}$ part of the metric, or in other words, what we view as the electromagnetic field, these light cones ``shift". The effect of this is that if we are considering three circles, $p \prec q \prec r$, then the overlap of the light cones of $p$ and $r$ on the circle $q$ will be reduced due to that relative shift. This shift is given by $f(p,q)$. Again, we have to be careful about path integration and, since $g_{\mu d} = f(p,q)/\tau(p,q)$, we have to use $\dd f(p,q)/\dd\tau(p,q)$ in our integration.

\bigskip
\noindent{\bf 5. Concluding Remarks}

\noindent In this paper we have demonstrated how gauge theory, coupled to a charged scalar field and gravity, can be translated into causal set terms. I should emphasize that the resulting discretized gauge theory is very different from the usual lattice gauge theory, as formulated on any type of lattice \cite{Rothe}.

I will conclude by adding some remarks on the relationship between the two approaches to gauge theory we have used, the conventional one and the Kaluza-Klein approach. In both cases we were using two-point functions on a causal set. In the case of the Kaluza-Klein model the two-point function is used as a replacement for causal relations in order to accommodate the situation of two causally related points so close to each other's light cone that parts of the circle in the extra dimension are not causally related. It is interesting to note that in the case of gauge theory the behavior of the two-point function inside the entire Alexandrov set is relevant to the Lagrangian, whereas in the case of Kaluza-Klein theory we have to be sufficiently close to the boundary of the Alexandrov set, since that is the only region where the behavior of $g_{\mu 5}$ is relevant to the causal structure; or, in the language of causal sets, that is the only region where the function that determines causal relations is neither 0 nor 1.

It is also interesting to relate these two theories: physically, the idea behind the Kaluza-Klein model can be explained by saying that the phase shift when you go around the loop in gauge theory is really a curvature effect, which arises when going around a loop in the 5th dimension. If we make this observation, then the fact that in both cases we are using two-point functions is not surprising: the two-point function might really be one and the same thing. It is both responsible for the ``shift" on our circles, thus affecting the volume of the Alexandrov set, and it is also responsible for the phase shift if we go around the loop picking the same points on the circle without ``shifting".
 
Writing the propagators while both gauge field and gravitational fields are allowed to vary is very problematic, because we are no longer allowed to use the causal structure in order to tell how far the points are separated, which means that we would have no information about any selected pair of points to compute the propagator. However, if the causal structure is fixed, we {\em can\/} use the results of the paper to compute propagators, which would be a causal set version of quantum field theory on a fixed curved background. In the case of the gauge theory model, we can integrate over holonomies, given a fixed information about the metric. In the case of the Kaluza-Klein model, we can integrate over our two-point function $d(p,q)$ under the constraint that $d(p,q) = 0$ whenever $p$ and $q$ are not causally related, where the causal structure is fixed. This is quite interesting, as it allows us to quantize Kaluza-Klein theory without quantizing gravity. Of course, a similar thing can be done in a regular, coordinate-based, Kaluza-Klein model; but in that case we face renormalization issues which are not relevant to causal sets.

As far as the complete theory coupled to gravity, which includes the variation of the metric, while there are no conclusive results by any means, some of the attempts to address that issue, including the issue of manifoldlike-ness, have been made in the paper on fermions in this series \cite{spinors}.

\bigskip
\noindent{\bf Acknowledgements}

\noindent I would like to express my great gratitude to Professor Luca Bombelli for discussing the content of this paper and for helping me edit it.


\end{document}